\documentclass[aps,preprint]{revtex4}%
\usepackage{amsfonts}
\usepackage{amsmath}
\usepackage{amssymb}
\usepackage{graphicx}%
\providecommand{\U}[1]{\protect \rule{.1in}{.1in}}
\begin{document}
\preprint{ }
\title[Short title for running header]{Solution of the Gross-Pitaevskii equation in terms of the associated
non-linear Hartree potential}
\author{George Rawitscher}
\affiliation{Physics Department, University of Connecticut}
\keywords{one two three}
\pacs{PACS number}

\begin{abstract}
The Gross-Pitaevskii equation (GP), that describes the wave function of a
number of coherent Bose particles contained in a trap, contains the cube of
the normalized wave function, times a factor proportional to the number of
coherent atoms. The square of the wave function, times the above mentioned
factor, is defned as the Hartree potential. A method implemented here for the
numerical solution of the GP equation consists in obtaining the Hartree
potential iteratively, starting with the Thomas Fermi approximation to this
potential. The energy eigenvalues and the corresponding wave functions for
each successive potential are obtained by a method described previously. After
approximately $35$ iterations a stability of eight significant figures for the
energy eigenvalues is obtained$.$This method has the advantage of being
physically intuitive, and could be extended to the calculation of a
shell-model potential in nuclear physics, once the Pauli exclusion principle
is allowed for.

\end{abstract}
\startpage{1}
\endpage{102}
\maketitle

\section{Introduction}

The phenomenon of Bose-Einstein condensation of an assembly of atoms,
predicted in 1924 \cite{EIN}, was finally observed experimentally in 1995
\cite{OB} for atoms confined in a trap at very low temperatures. An
approximate non-linear equation that describes the BEC was established in 1961
by E. P.Gross \cite{GRO}, and independently by L. P. Pitaevskii \cite{PIT}.
This is a Schr\"{o}dinger-like equation, now called the Gross-Pitaevskii
equation (GPE), for the wave function of $N$ Bose particles interacting
coherently confined in an atomic trap. In this equation only the short range
part of the interaction between the atoms is included in terms of the
scattering length of two colliding atoms. That term is proportional to the
cube of the wave function, with a coefficient that is proportional to $N$ and
to the scattering length $a$. Numerical solutions of this non-linear GPE began
to be obtained in the middle 60'ies, both for the time independent form
\cite{EB}, as well as for the time dependent form \cite{RHB}. An extensive
review of the early work is given in Ref. \cite{RMP}, that contains more than
240 references. Both the experimental as well as theoretical work continues
actively today. On the theoretical side various\ diverse methods for the
solution of the GP equation have been developed. Amongst them, some based on
mathematical theorems \cite{CKMS}, others based on spectral expansions
\cite{DC}, others using extensive numerical methods \cite{SF}, and others that
also include the interaction of the BEC atoms with the surrounding atomic
medium \cite{BFBS}. An article by Bao, Jaksh and Markowich \cite{BJM} contains
references to such studies.

One aspect emphasized in the present study is the description of the coherent
interaction of the atoms in the BEC in terms of the related Hartree potential,
$V_{H}$. This potential arises naturally in the GPE, due to the presence of
the third power of the wave function $\Psi$ in that equation, by rewriting the
term $\varpropto \Psi^{3}$ as $V_{H}\Psi$. This potential contains the square
of the wave function, and hence is nonlinear. Such a term was introduced in
the context of fluid dynamics by E. P. Gross \cite{GROSS}, and in the context
of nuclear physics it is called the Hartree-Fock potential since it
incorporates the effect of the Pauli exclusion imposed on the fermions
\cite{NEG}.

If $V_{H}$ were known, then the GP equation could be written as an ordinary
linear $Schr\ddot{o}dinger$ equation, that could be solved by conventional
means for the ground or excited states of $\Psi$. Since $V_{H}$ is not known,
it can nevertheless be solved for iteratively, by starting from a good
approximation to $V_{H},$ solving for the corresponding wave function that in
turn defines a better approximation to $V_{H}$, and so on. To demonstrate the
viability of this scheme is the purpose of the present paper.

It is found for the present numerical examples that the iterations converge,
and since the convergence is non-monotonic, it is expected that the converged
solution becomes unique. However, no attempt was made in the present study to
determine the upper value of the number of coherent atoms $N$ in the trap
beyond which the iterations diverge. Since no variational methods are involved
in the calculation, both the ground and several excited states of the BEC can
be found without much difficulty. As a function of the radial distance Hartree
potentials are monotonic for the ground state, and oscillatory in different
ways for the excited states. One conclusion is that no one single mean-field
potential is able to give rise simultaneously to the ground and the various
excited BEC states. A future envisaged application of this method is \ in
\ the calculation of a shell model potential in nuclear physics. In this case
several (but not many) nucleons occupy a given "shell", but the confining
potential will turn out to be different for each shell. Hence the shell
potential becomes non-local, and it is hoped that the present method may
facilitate the formulation of this non-locality. Similarly, the optical model
potential describing nucleon-nucleus scattering is also non-local, (but for
more reasons) and efforts to determine its nature are in progress \cite{OPTM}.

The present investigation is limited to a spherically symmetric confining
well, and only the partial wave corresponding to an angular momentum $L=0$ is
included. The confining well is assumed to be harmonic, but other forms can
also be considered. The organization of this paper is as follows: In Section
II the formalism of the GP equation is reviewed, a physically justified set of
input parameters is proposed, and the Thomas-Fermi approximation to $V_{H}$ is
implemented. Section III contains results for $V_{H}$ and the corresponding
excitation energies, and section IV contains the summary and conclusions.

\bigskip

\section{Formalism}

The three-dimensional form of the (GP) equation can be written \cite{RMP}%
\begin{equation}
i\hbar \frac{\partial}{\partial t}\Psi(\vec{r},t)=\left[  -\frac{\hbar^{2}}%
{2m}\nabla^{2}+V_{ext}(\vec{r})+g|\Psi|^{2}\right]  \Psi(\vec{r},t),
\label{GP3}%
\end{equation}
where $\hbar$ is Planck's constant divided by $2\pi$, $m$ is the mass of the
Boson, $V_{ext}$ is the confining trap potential, usually written as a sum of
three harmonic potentials $\omega_{x}x^{2}+\omega_{y}y^{2}+\omega_{z}z^{2}$,
and $g$ is a constant proportional to the number $N$ of particles in the trap
times the scattering length $a$ of two of the Bosons. This constant can be
written as \cite{EB}
\begin{equation}
g=NU_{0}, \label{g}%
\end{equation}
with
\begin{equation}
U_{0}=\frac{\hbar^{2}}{2m}8\pi a. \label{U0}%
\end{equation}
A stationary solution $\Psi(\vec{r},t)=\exp(-i\mu t/\hbar)\psi(\vec{r})$ obeys
\cite{EB}
\begin{equation}
\mu \psi(\vec{r})=\left[  -\frac{\hbar^{2}}{2m}\nabla^{2}+V_{ext}(\vec
{r})+NU_{0}|\psi(\vec{r})|^{2}\right]  \psi(\vec{r}). \label{GP3stat}%
\end{equation}

If one now assumes that $V_{ext}=V(r),$ i.e., that the trap potential is
spherically symmetric, makes a partial wave expansion of $\psi(\vec{r})$, and
retains only the angular momentum $L=0$ part of the expansion,
\begin{equation}
\psi(\vec{r})\rightarrow \psi(r)=\phi(r)/(r\sqrt{4\pi}), \label{L0}%
\end{equation}
then $\phi(r)$ satisfies the radial equation \cite{EB}
\begin{equation}
\left[  -\frac{\hbar^{2}}{2m}\frac{d^{2}}{dr^{2}}+V_{ext}(r)+N\frac{U_{0}%
}{4\pi}|\phi(r)/r|^{2}\right]  \phi(r)=\mu \phi(r). \label{GP1stat}%
\end{equation}
Here $\psi(\vec{r})$ describes the wave function of one of the particles, and
since the probability of finding this particle is unity, i.e., $\int|\psi
(\vec{r})|^{2}d^{3}\vec{r}=1,$ one finds, in view of Eq. (\ref{L0}),%
\begin{equation}
\int_{0}^{\infty}|\phi(r)|^{2}dr=1. \label{NORM}%
\end{equation}

The first, second, etc., iterations of $\phi(r)$ are denoted as $\phi
^{(1)},\phi^{(2)},...\phi^{(n)}..$, the corresponding Hartree potentials are
denoted as
\begin{equation}
V_{H}^{(n)}(r)=N\frac{U_{0}}{4\pi}|\phi^{(n)}(r)/r|^{2},~~n=0,1,2,...,
\label{VHn}%
\end{equation}
and the iterative equations are
\begin{equation}
\left[  -\frac{\hbar^{2}}{2m}\frac{d^{2}}{dr^{2}}+V_{ext}(r)+V_{H}%
^{(n)}(r)\right]  \phi^{(n+1)}(r)=\mu^{(n+1)}\phi^{(n+1)}(r),\  \  \ n=0,1,2,..
\label{PHIit}%
\end{equation}
The functions $\phi^{(n)}$ all go to zero at the origin of $r$, decay to zero
as gaussians as $r\rightarrow \infty$ if $V_{ext}$ is assumed to be harmonic,
and obey the normalization condition (\ref{NORM}) for each iteration. For each
fixed value of $(n)$, the eigenfunctions $\phi^{(n+1)}$ and eigenvalues
$\mu^{(n+1)}$ of Eqs. (\ref{PHIit}) are determined iteratively by a Hartree
procedure described previously, both for bound states \cite{HEHE} as well as
for Sturmian eigenvalues \cite{GRSTU}.

In summary, two nested iterations are performed: 1. One that finds the
solutions of Eq.(\ref{PHIit}) for each value of $(n)$, and 2. The iterative
progression from $(n)$ to $(n+1).$ The latter proceeds non-monotonically, as
seen in the numerical example given further on, and the first has been used
successfully in several applications \cite{SPEC-C}. This double iteration
procedure is different from the procedures cited above \cite{EB},\cite{RHB},
\cite{CKMS}-\cite{BJM}, \  \cite{ADH}. Another difference from previous
calculations is that the differential equation (\ref{PHIit}) is transformed
into a Lippmann-Schwinger integral integral equation, that is solved with the
use of Green's functions in configuration space. These functions require
wave-numbers, rather than energies as input parameters. The calculations are
done by means of a semi-spectral Chebyshev expansion method that gives a
reliable accuracy \cite{SPEC-A}, \cite{SPEC-B}.

\subsection{Numerical inputs}

In order to solve Eqs. (\ref{VHn}) and (\ref{PHIit}), two steps are required.
First a set of physically reasonable values for the potentials have to be
established, and subsequently a transformation of variables is made so as to
render the equations more transparent, and all quantities become expressed in
terms of new distance and energy units.

The atoms in the trap are assumed to have a mass $m=30u,$ and the scattering
length $a=3nm.$ The confining trap potential is assumed to be harmonic
\begin{equation}
V_{ext}=\alpha r^{2}, \label{Vext}%
\end{equation}
and the value of the coefficient $\alpha$ is obtained by requiring that at a
distance of $1\mu m$ from the center of the trap the value of $V_{ext}%
=100\ kT,$ with $T=10^{-9}K.$ This yields $\alpha=8.5\ eV/m^{2}.$ Next, both
sides of Eq. (\ref{GP1stat}) are multiplied by $2m/\hbar^{2}$, a new unit of
distance $D$ is chosen
\begin{equation}
D=(\frac{\hbar^{2}}{2m}\frac{1}{4\alpha})^{(1/4)}\simeq3.8\times10^{-7}%
m\simeq7000\ a_{Bohr} \label{D}%
\end{equation}
and by further multiplying by $D^{2},$ Eq. (\ref{PHIit}) is transformed into
dimensionless units%
\begin{equation}
-\frac{d^{2}\phi}{dx^{2}}+\left[  \frac{1}{4}x^{2}+N\beta|\phi(x)/x|^{2}%
\right]  \phi(x)=\lambda \phi(x), \label{GP1x}%
\end{equation}
and the normalization Eq. (\ref{NORM}) is changed to
\begin{equation}
\int_{0}^{\infty}|\phi(x)|^{2}dx=1 \label{NORMx}%
\end{equation}
Here
\begin{equation}
x=r/D, \label{roD}%
\end{equation}%
\[
\beta=2a/D\simeq0.016,
\]%
\begin{equation}
\lambda=\frac{2m}{\hbar^{2}}D^{2}\mu=\mu/\varepsilon_{0}. \label{lambda}%
\end{equation}
The energy unit $\varepsilon_{0}$ is thus
\begin{equation}
\varepsilon_{0}=\frac{\hbar^{2}}{2m}\frac{1}{D^{2}}=(4\alpha \frac{\hbar^{2}%
}{2m})^{1/2}\approx5\times10^{-12}eV \label{EU}%
\end{equation}

In order to solve Eq. (\ref{GP1x}) numerically, a constant $V_{0}$ is
subtracted from both sides%
\[
V_{0}=20
\]
with the result%
\begin{equation}
-\frac{d^{2}\phi}{dx^{2}}+\left[  V(x)\right]  \phi(x)=-\kappa^{2}\phi(x),
\label{GPnum}%
\end{equation}
where
\begin{equation}
V(x)=(-V_{0}+\frac{1}{4}x^{2})+V_{H}(x), \label{Vof x}%
\end{equation}%
\begin{equation}
-\kappa^{2}=\lambda-V_{0}, \label{kappa2}%
\end{equation}
and where the dimensionless Hartee potential is given by%
\begin{equation}
V_{H}(x)=N\times \ 0.016\  \times|\phi(x)/x|^{2}. \label{VHx}%
\end{equation}

The effect of $V_{0}$ is to move the bottom of the harmonic well to a negative
energy, but $\lambda$ still measures the eigenvalue energy above the bottom of
the well. To this, thus moved harmonic potential is added the Hartree
potential $N\beta \ |\phi(x)/x|^{2}$, which is positive (repulsive) if the
scattering length $a$ is positive. The advantage of having subtracted $V_{0}$
is that the wave number $k$ required as input to the Green's function
$\mathcal{G}(k,x,x^{\prime})$ becomes purely imaginary, $k=i\kappa,$ and thus
the asymptotic value of $\mathcal{G}$ decreases exponentially. However, since
the potential $V$ continues to grow positively as $x$ increases, the
asymptotic form of $\phi(x)$ should decrease to zero like a Gaussian function.
This behavior is indeed found to be the case in the numerical evaluations.

\subsection{The Thomas-Fermi approximation}

This approximation to $V_{H}$ is obtained by dropping the kinetic energy term
from the GP equations (\ref{GP3stat}) or (\ref{GP1stat}). As already noted
previously \cite{TF}, this approximation, denoted as $V_{TF},$ gets better the
larger the number $N$ of coherent atoms in the trap. However, since the
function $V_{TF}$ drops abruptly to zero at the outer edge of $V_{TF}$, it is
difficult to incorporate this function into the numerical calculations
\cite{SF}. This difficulty is overcome in the present investigation, by
fitting to $V_{TF}$ \ a smooth extension that decreases to zero exponentially,
and subsequently using this fit for the start of the iterations for $V_{H}$.
The derivation of $V_{TF}$ will be repeated here for completeness.

\begin{figure}
[ptb]
\begin{center}
\includegraphics[
height=1.9164in,
width=2.5477in
]%
{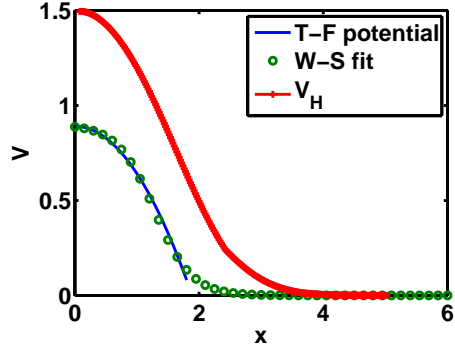}%
\caption{The Thomas Fermi aproximation to $V_{H}$ for $N=250$ and
$\beta=0.016$ is represented by the solid line. The Woods-Saxon fit to this
potential is represented by open circles, while the final non linear Hartree
potential for the ground state is represented by the thick line.}%
\label{FIG3}%
\end{center}
\end{figure}

By discarding the second order derivative in Eq. (\ref{GP1x}), one obtains%
\begin{equation}
N\beta|\phi|^{2}=x^{2}(\lambda-\frac{1}{4}x^{2}), \label{TF1}%
\end{equation}
where the maximum value of $x$ is $x_{M}=(4\lambda)^{1/2}.$ The value of
$\lambda$ is not known until one takes into account the normalization
condition (\ref{NORMx}). The integrals can be done analytically for the case
that the confining potential is harmonic, with the result%
\begin{equation}
\lambda_{TF}=\left[  \frac{15}{16}N\beta \right]  ^{(2/5)}. \label{lambdaTF}%
\end{equation}
The corresponding value of $x_{M}$ is%
\begin{equation}
x_{M}=2\left[  \frac{15}{16}N\beta \right]  ^{(1/5} \label{xM}%
\end{equation}
A numerical example for the case that $N=250$ is illustrated in Fig.
\ref{FIG3}.%

%TCIMACRO{\FRAME{ftbpFU}{2.5477in}{1.9164in}{0pt}{\Qcb{The Thomas Fermi
%aproximation to $V_{H}$ for $N=250$ and $\beta=0.016$ is represented by the
%solid line. The Woods-Saxon fit to this potential is represented by open
%circles, while the final non linear Hartree potential for the ground state is
%represented by the thick line.}}{\Qlb{FIG3}}{tf_hf_pot_4.eps}%
%width 2.5477in;  height 1.9164in;  depth 0pt;  original-width 5.8219in;
%original-height 4.3708in;  cropleft "0";  croptop "1";  cropright "1";
%cropbottom "0";  filename 'TF_HF_pot_4.eps';file-properties "XNPEU";}} }%
%BeginExpansion
%EndExpansion

\section{Results}

As described in Section II the calculation consists of two nested iterations.
For each Hartree potential $V_{H}^{(n)}$ the corresponding eigenvalue
$\lambda^{(n+1)}$ and eigenfunction $\phi^{(n+1)}(x)$ is calculated by a
hybrid\ iterative method, implemented by means of a spectral Chebyshev
expansion described in Ref. \cite{HEHE}. The resulting Hartree potential
$V_{H}^{(n+1)}$, given by $N\beta|\phi^{(n+1)}(x)/x|^{2},$ is thus obtained,
and so forth. Two different methods are used in order to initiate the procedure.

\begin{figure}
[ptb]
\begin{center}
\includegraphics[
height=2.2243in,
width=2.9577in
]%
{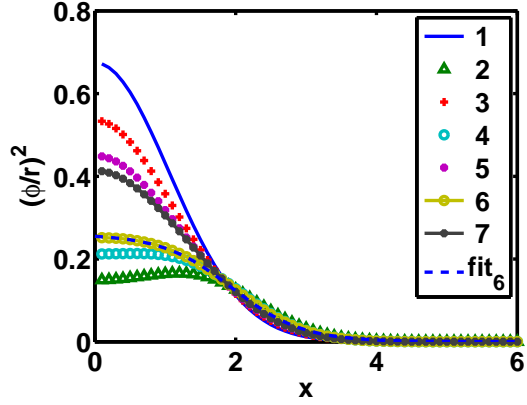}%
\caption{Iterative values of $(\phi^{(n)}(x)/x)^{2}$ as a function of the
dimensionless radial distance $x=r/D$, for the ground BEC state. The iteration
number $n$ is shown in the legend. The iterations start with the ground-state
solution $\phi$ of the harmonic potential $-20+0.2x^{2}$, that, in view of Eq.
(\ref{VHx}) with $N=250$, provides the first value to $V_{H}$ and hence of
$V(x),$ defined in Eq. (\ref{VHx}). The functions\ $(\phi^{(n)}(x)/x)^{2}$ for
$n=1...7,$ are fitted by hand with a combination of Wood-Saxon functions in
order to obtain an approximation to the next Hartree potential. \ For each
iteration the normalization of $\phi^{(n)}$ is given by $\int_{0}^{\infty
}(\phi^{(n)})^{2}dx=1$ \ }%
\label{FIG1}%
\end{center}
\end{figure}

\begin{figure}
[ptb]
\begin{center}
\includegraphics[
height=2.1767in,
width=2.8971in
]%
{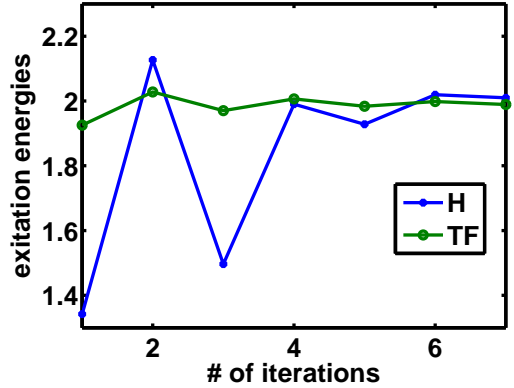}%
\caption{The ground state energies above the bottom of the attractive trap
well, as a function of the number $n$ $\ $of iterations. The iterations
labeled "H" were started with the eigenfunction of the Harmonic well, while
the ones labelled "TF" where started with a fit to the Thomas-Fermi
approximation to the Hartree potentia. The conditions are the same as in
Fig.\  \ref{FIG1}.}%
\label{FIG2}%
\end{center}
\end{figure}

The first starts from the eigenfunction of the harmonic potential, in the
absence of $V_{H},$ the resulting function $|\phi^{(1)}(x)/x|^{2}$ is fitted
with a Woods-Saxon form, and after multiplication by $N\beta$ the value of
$V_{H}^{(1)}$ is obtained, and the process is repeated for subsequent
iterations. Results with this method for the values $|\phi^{(n)}(x)/x|^{2}$
are illustrated in Fig. \ref{FIG1} for the ground state solutions. The
convergence is oscillatory, and the gap between successive values of
\ $|\phi^{(n)}(x)/x|^{2}$ gradually decreases. The corresponding values of the
ground state excitation energy are illustrated in Fig. \ref{FIG2} by the
points labelled as "H", which also shows the oscillatory nature of the
convergence. The first point, close to 1.4, corresponds to the excitation
energy for the pure harmonic oscillator, which is smaller that the final
excitation energy, close to 2.0, that is due to the repulsive nature of the
Hartree potential.
%TCIMACRO{\FRAME{ftbpFU}{2.9577in}{2.2243in}{0pt}{\Qcb{Iterative values of
%$(\phi^{(n)}(x)/x)^{2}$ as a function of the dimensionless radial distance
%$x=r/D$, for the ground BEC state. The iteration number $n$ is shown in the
%legend. The iterations start with the ground-state solution $\phi$ of the
%harmonic potential $-20+0.2x^{2}$, that, in view of Eq. (\ref{VHx}) with
%$N=250$, provides the first value to $V_{H}$ and hence of $V(x),$ defined in
%Eq. (\ref{VHx}). The functions\ $(\phi^{(n)}(x)/x)^{2}$ for $n=1...7,$ are
%fitted by hand with a combination of Wood-Saxon functions in order to obtain
%an approximation to the next Hartree potential. \ For each iteration the
%normalization of $\phi^{(n)}$ is given by $\int_{0}^{\infty}(\phi^{(n)}%
%)^{2}dx=1$ \ }}{\Qlb{FIG1}}{rpsi2_4.eps}%
%width 2.9577in;  height 2.2243in;  depth 0pt;  original-width 5.8219in;
%original-height 4.3708in;  cropleft "0";  croptop "1";  cropright "1";
%cropbottom "0";  filename 'rpsi2_4.eps';file-properties "XNPEU";}} }%
%BeginExpansion
%EndExpansion
%

%TCIMACRO{\FRAME{ftbpFU}{2.8971in}{2.1767in}{0pt}{\Qcb{The ground state
%energies above the bottom of the attractive trap well, as a function of the
%number $n$ $\ $of iterations. The iterations labeled "H" were started with the
%eigenfunction of the Harmonic well, while the ones labelled "TF" where started
%with a fit to the Thomas-Fermi approximation to the Hartree potentia. The
%conditions are the same as in Fig.\ \ref{FIG1}.}}{\Qlb{FIG2}}{exit_4.eps}%
%width 2.8971in;  height 2.1767in;  depth 0pt;  original-width 5.8219in;
%original-height 4.3708in;  cropleft "0";  croptop "1";  cropright "1";
%cropbottom "0";  filename 'exit_4.eps';file-properties "XNPEU";}} }%
%BeginExpansion
%EndExpansion
The second method starts the iteration with a smoothed fit to the Thomas Fermi
potential, as shown by the open circles in Fig. \ref{FIG3}. The corresponding
excitation energies are displayed by the open circles in Fig. \ref{FIG2}. It
is clear that the Thomas Fermi form for the Hartree potential provides a much
better starting approximation for the iterations than the harmonic oscillator
eigenfunction.
%TCIMACRO{\FRAME{ftbpFU}{2.6731in}{2.0098in}{0pt}{\Qcb{The final energies, in
%units of $\varepsilon_{0},$ of the ground, first and second excited states.
%The lowest set of points correspond to the harmonic well alone, and the other
%points are for the GP cases with $N=250$ and $1000$, respectively. The ground,
%first and second excited states are located on the $x-$axis at the points $0,$
%$1,$ and $2$, respectively. }}{\Qlb{FIG4}}{exc_energ.eps}%
%width 2.6731in;  height 2.0098in;  depth 0pt;  original-width 5.8219in;
%original-height 4.3708in;  cropleft "0";  croptop "1";  cropright "1";
%cropbottom "0";  filename 'exc_energ.eps';file-properties "XNPEU";}} }%
%BeginExpansion
\begin{figure}
[ptb]
\begin{center}
\includegraphics[
height=2.0098in,
width=2.6731in
]%
{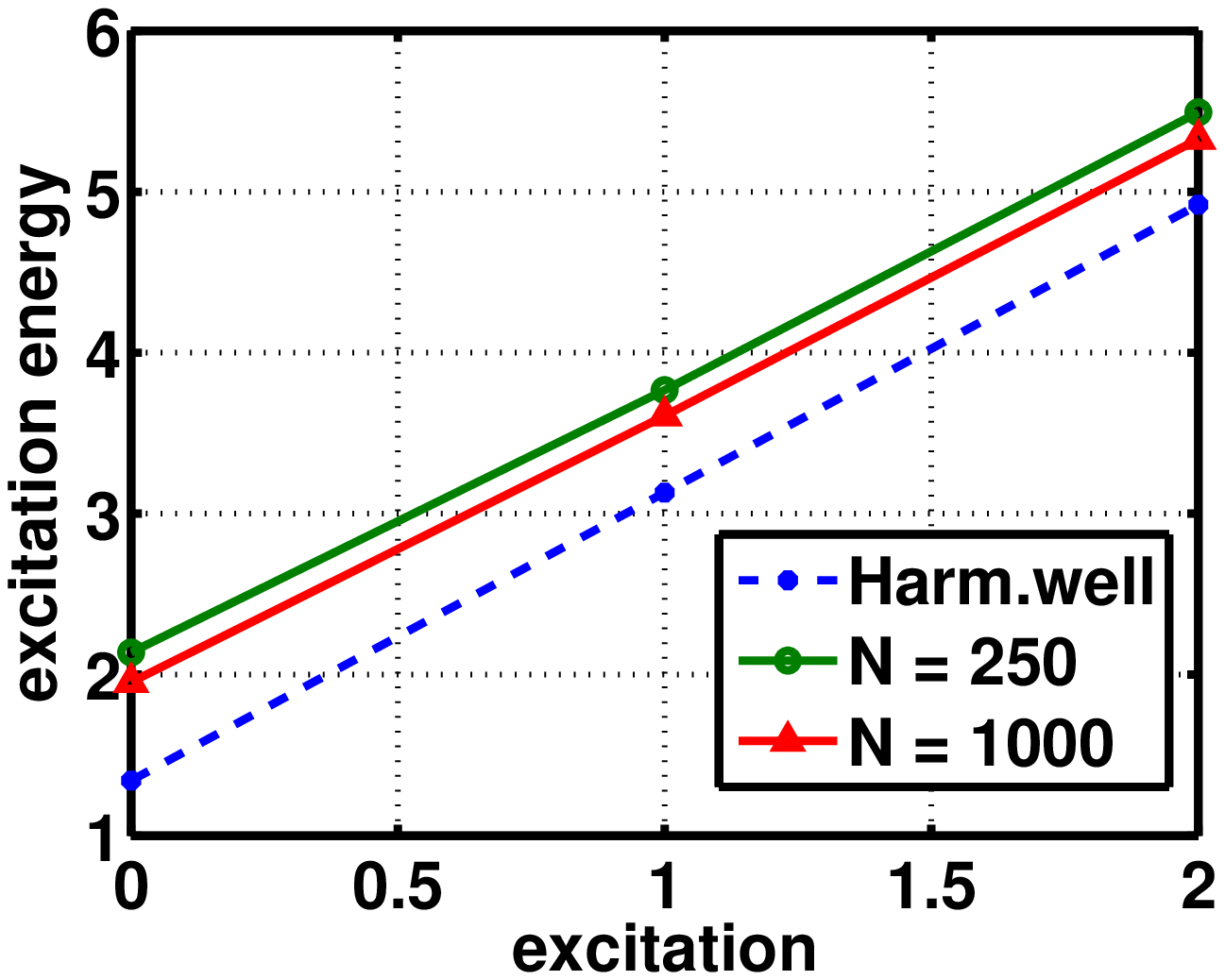}%
\caption{The final energies, in units of $\varepsilon_{0},$ of the ground,
first and second excited states. The lowest set of points correspond to the
harmonic well alone, and the other points are for the GP cases with $N=250$
and $1000$, respectively. The ground, first and second excited states are
located on the $x-$axis at the points $0,$ $1,$ and $2$, respectively. }%
\label{FIG4}%
\end{center}
\end{figure}
%EndExpansion

Not only the ground state of the GP equation can be obtained with this
iterative method starting from the fitted Thomas Fermi (TF) approximation to
the Hartree potential, but with the same TF potential the higher excited
states can also be obtained iteratively. The results for the ground, first and
second excited states are displayed in Fig. \ref{FIG4}. The excitation
energies for the GP equation lie above the values for the pure Harmonic
potential well, confirming that the corresponding Hartree potentials are
repulsive. It is interesting to note that for a larger value of the number $N$
of coherent particles, the excitation energies are slightly lower. According
to Eq. (\ref{lambda}) these energies are given in units of $\varepsilon_{0}$,
Eq. (\ref{EU})$.$
%TCIMACRO{\FRAME{ftbpFU}{2.6662in}{2.0046in}{0pt}{\Qcb{The sum of Harmonic and
%converged Hartree potentials for the ground, first and second BEC excited
%states, for $N=250.$.}}{\Qlb{FIG9}}{vh_n4_n0_1_2.eps}%
%width 2.6662in;  height 2.0046in;  depth 0pt;  original-width 5.8219in;
%original-height 4.3708in;  cropleft "0";  croptop "1";  cropright "1";
%cropbottom "0";  filename 'vH_N4_n0_1_2.eps';file-properties "XNPEU";}} }%
%BeginExpansion
\begin{figure}
[ptb]
\begin{center}
\includegraphics[
height=2.0046in,
width=2.6662in
]%
{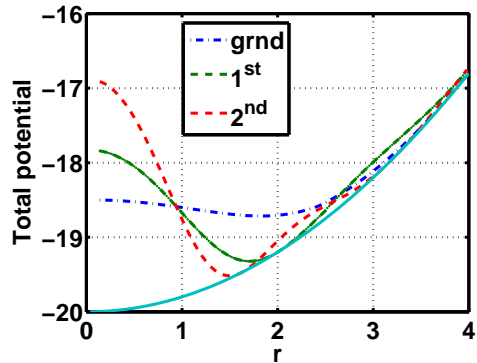}%
\caption{The sum of Harmonic and converged Hartree potentials for the ground,
first and second BEC excited states, for $N=250.$.}%
\label{FIG9}%
\end{center}
\end{figure}
%EndExpansion
The Hartree potentials, when added to the harmonic trap potential, are
displayed in Figs. \ref{FIG9} and \ref{FIG8} for the values of $N=250$ and
$1000$, respectively. The properties of the Hartree potentials can be inferred
from these graphs: as $N$ increases, these potentials increase proportionally,
but the functions $|\phi(x)/x|^{2}$ do not change significantly.%
%TCIMACRO{\FRAME{ftbpFU}{2.6515in}{1.9951in}{0pt}{\Qcb{Same as Fig. \ref{FIG9}
%for $N=1000.$ Please note the change in scale of the $y-$ axix.}}{\Qlb{FIG8}%
%}{vh_n16_n0_1_2.eps}{\special{ language "Scientific Word";  type "GRAPHIC";
%width 2.6515in;  height 1.9951in;  depth 0pt;  original-width 5.8219in;
%original-height 4.3708in;  cropleft "0";  croptop "1";  cropright "1";
%cropbottom "0";  filename 'vH_N16_n0_1_2.eps';file-properties "XNPEU";}} }%
%BeginExpansion
\begin{figure}
[ptb]
\begin{center}
\includegraphics[
height=1.9951in,
width=2.6515in
]%
{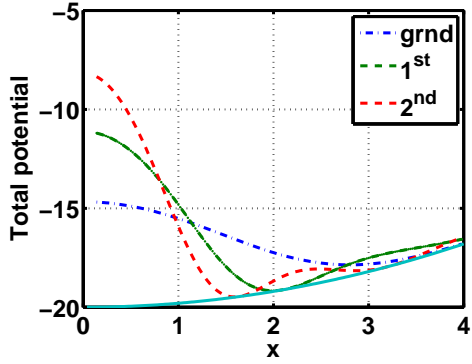}%
\caption{Same as Fig. \ref{FIG9} for $N=1000.$ Please note the change in scale
of the $y-$ axix.}%
\label{FIG8}%
\end{center}
\end{figure}
%EndExpansion

\subsection{Computational details}

The calculations are done with MATLAB on a desk PC using an Intel TM2 Quad,
with a CPU Q 9950, a frequency of 2.83 GHz, and a RAM\ of 8 GB. For the case
of $N=250,$ forty iterations take between $6$ and $7$ seconds. Table
\ref{TABLE2}\ of energy values for the ground and first excited states (in
units of $\varepsilon_{0})$ indicates the rate of convergence.%

%TCIMACRO{\TeXButton{B}{\begin{table}[tbp] \centering}}%
%BeginExpansion
\begin{table}[tbp] \centering
%EndExpansion%
\begin{tabular}
[c]{|l|l|l|l|}\hline
Iteration & grnd. st. & 1'st exc. st. & grnd. st.\\ \hline \hline
$\#$ & $N=250$ & $N=250$ & $N=1000$\\ \hline \hline
$1$ & $2.1$ & $4$ & $2$\\ \hline
$5$ & $2.14$ & $3.8$ & $2$\\ \hline
$10$ & $2.14$ & $3.77$ & $1.95$\\ \hline
$15$ & $2.139$ & $3.768$ & $1.952$\\ \hline
$20$ & $2.1388$ & $3.768$ & $1.9525$\\ \hline
$25$ & $2.13882$ & $3.76775$ & $1.95250$\\ \hline
$30$ & $2.138821$ & $3.767749$ & $1.952498$\\ \hline
$35$ & $2.1388211$ & $3.7677496$ & $1.9524984$\\ \hline
\end{tabular}
\caption{Convergence of the excitation energies in units described in the text}\label{TABLE2}%
%TCIMACRO{\TeXButton{E}{\end{table}}}%
%BeginExpansion
\end{table}%
%EndExpansion

\section{Summary and Conclusions}

A method is presented of solving for the $L=0$ partial wave-function of the
the Gross Pitaevskii (GP) nonlinear differential equation, that approximates
the wave function for atoms that are bound in a spherically symmetric harmonic
oscillator trap potential. A Hartree potential $V_{H}$ is used as a key
vehicle for performing the iterations that converge for a low number of $N$ of
atoms in the trap. This potential is defined as the wave-function squared
times a factor proportional to the number $N$ of coherent atoms and the
(positive) scattering length. The parameters of the equation are determined
from physical considerations. The Hartree potentials and binding energies are
obtained for the ground, first and second excited states for $N=250$ and
$1000.$ It is found that the start of the iterative process based on the
Thomas-Fermi approximation to $V_{H}$ \ is more efficient than when the
iterations are started from the eigenfunction of the harmonic well, as is
shown in Fig. \ref{FIG2}. The iterations that lead from one $V_{H}$ to the
next, as described in Eq. (\ref{PHIit}), converge rather slowly. After each 5
iterations the stability of the excitation energy increases approximately by
one significant figure, but the computational complexity is not excessive. The
knowledge of $V_{H}$ is suggestive for future applications, such as for
refining a mean-field potential for nucleons in a nucleus, once the Pauli
exclusion principle for the nucleons is taken into account. This approach may
lead to different nuclear mean field potentials for different shells.


\begin{thebibliography}{99}                                                                                               %


\bibitem {EIN}S. N. Bose, Z. Phys 26 (1924)178; A. Einstein, Sitzber. Kg.
Preuss. Akad. Wiss, 261 (1924);A. Einstein, ibid. 3, (1925);

\bibitem {OB}M. H. Anserson, J. R. Ensher, M. R. Matthews, C. E. Wieman, and
E. A. Cornell, Science \textbf{269, }198\textbf{ (}1995); K. B. Davis, M. O.
Mewes, M. R. Andrews, N. J. van Druten, D. S. Durfee, D. M. Kurn, and W.
Ketterle, Phys. Rev. Lett. \textbf{75},3969 (1995);

\bibitem {GRO}E. P. Gross, Nuovo Cimento \textbf{20}, 454 (1961); J. Math
Phys. \textbf{4}, 195 (1963);

\bibitem {PIT}L. P. Pitaevskii, Zh Eksp. Theor. Fiz \textbf{40, }646
(1961)\textbf{ [}Sov. Phys.\textbf{ }JETP\textbf{ 13}, 451 (1961)]; Phys.
Lett. A \textbf{221} (1996);

\bibitem {EB}M. Edwards, and K. Burnett, Phys. Rev. A \textbf{51},1382 ( 1995);

\bibitem {RHB}P. A. Ruprecht, M. J. Holland, and K. Burnett, M. Edwards, Phys.
Rev. A \textbf{51},4704 (1995);

\bibitem {RMP}F. Dalfovo, S. Giorgini, L. P. Pitaevskii, and S. Stingari, Rev.
Mod. Phys. \textbf{71}, 463 (1999);

\bibitem {CKMS}Y. -S. Choi, I. Koltracht, P. J. McKenna, and N. Savytska,
Linear Algebra and its applications \textbf{357}, 217 (2002); Y-S Choi, J.
Javanainen, I. Koltracht, M. Ko\v{s}trun, P. J. McKenna, and N. Savytska, J.
of Comp. Phys.\textbf{190}, 1 (2003);

\bibitem {DC}C. M. Dion and E. Canc\`{e}s, Phys. Rev. E \textbf{67}, 046706 (2003);

\bibitem {SF}B. I. Schneider and D. L. Feder, \ Pys. Rev. A \textbf{59}, 2232 (1999);

\bibitem {BFBS}T. Bergeman, D. L. Feder, N. L. Balazs, and B. I. Schneider,
Phys. Rev. A \textbf{61}, 063605 (2000); A. Gammal,T. Frederico, L. Thomio, Ph
Chomaz, J. Phys B \textbf{33}, 4053 (2000); W. Jiang, H. Wang and X. Li, Comp.
Phys. Comm, in press (2013);

\bibitem {BJM}W. Bao, D. Jaksch, and P. A. Markowich, arXiv:\ cond-mat/0303239
v1 (2003);

\bibitem {GROSS}E. P. Gross, J. Math. Phys. \textbf{4}, 195 (1963); J. C. Gunn
and J. M. F. Gunn, Eur. J. Phys. \textbf{9}, 51 (1988);

\bibitem {NEG}J. W. Negele, Phys. Rev. \textbf{C1}, 1260\ (1970); J. W. Negele
and D. Vautherin, Phys. Rev. \textbf{\ C 5}, 1472 (1971); D. Vautherin and D.
M. Brink, Phys. Rev. \textbf{C 5}, 626 (1972);

\bibitem {OPTM}{S. G. Cooper and R. S. Mackintosh, \textit{Phys. Rev.C}
\textbf{54,} 3133 (1996) ;} {K. Amos, L. Canton, G. Pisent, J. P. Svenne, D.
van der Knijff, \textit{Nucl. Phys. A} \textbf{728} 65 (2003); M. I. Jaghoub,
and G. H. Rawitscher, Nucl. Phys. \textbf{A 877}, 59 (2012);}

\bibitem {HEHE}G. Rawitscher and I. Koltracht, Eur. J. Phys. \textbf{27}, 1179
( 2006);

\bibitem {GRSTU}G. Rawitscher, Phys. Rev. E \textbf{85}, 026701 (2012);

\bibitem {SPEC-C}G. Rawitscher, \emph{Applications of a numerical spectral
expansion method to problems in physics: A retrospective, }in Operator Theory,
Advances and Applications, Vol. 203, edited by Thomas Hempfling (Birk\"{a}user
Verlag, Basel, 2009), pp. 409--426;

\bibitem {ADH}S. K. Adhikari, Phys. Rev. E, \textbf{62},2937 (2000), ibid
\textbf{63}, 054502 (2001);

\bibitem {SPEC-A}R. A. Gonzales, J. Eisert, I Koltracht, M. Neumann and G.
Rawitscher, J. of Comput. Phys. \textbf{134}, 134-149 (1997); R. A. Gonzales,
S.-Y. Kang, I. Koltracht and G. Rawitscher, J. of Comput. Phys. \textbf{153},
160-202 (1999);

\bibitem {SPEC-B}A. Deloff, Ann. Phys. (NY) 322, 1373--1419 (2007); L. N.
Trefethen, \emph{Spectral Methods in MATLAB}, (SIAM, Philadelphia, PA, 2000) ;
John P. Boyd, \emph{Chebyshev and Fourier Spectral Methods,} 2nd revised ed.
(Dover Publications, Mineola, NY, 2001); G. Rawitscher and I. Koltracht,
Computing Sci. Eng. \textbf{7}, 58 (2005);

\bibitem {TF}F. Dalfovo and S. Stringari, Phys. Rev. A \textbf{53}, 2477
(1996); G. Baym and C. J. Pethick, Phys. Rev. Lett \textbf{76}, 6 (1996);
\end{thebibliography}
\end{document}